\documentclass[floats, eqnum, showpacs, nofootinbib, twocolumn, eqsecnum]{revtex4-1}

\usepackage{color,graphicx}
\usepackage{amsfonts}
\usepackage{amssymb}

\begin{document}

\title{Linear stability analysis of a rotating thin-shell wormhole}
\author{Naoki Tsukamoto${}^{1}$}\email{tsukamoto@rikkyo.ac.jp}
\author{Takafumi Kokubu${}^{2,3}$}\email{14ra002a@rikkyo.ac.jp, kokubu@post.kek.jp}
\affiliation{
${}^{1}$Frontier Research Institute for Interdisciplinary Sciences \& Department of Physics, Tohoku University, 
Sendai, Miyagi 980-8578, Japan \\
${}^{2}$Theory Center, Institute of Particle and Nuclear Studies, \\
KEK (High Energy Accelerator Research Organization), Tsukuba, Ibaraki 305-0801, Japan\\
${}^{3}$Department of Physics, Rikkyo University, Toshima, Tokyo 171-8501, Japan
}

\begin{abstract}
We cut and paste two Ba\~{n}ados-Teitelboim-Zanelli (BTZ) spacetimes at a throat 
by the Darmois-Israel method to construct a rotating wormhole with a thin shell filled with a barotropic fluid.
The thin shell at the throat and both sides of the throat corotate.
We investigate the linear stability of the thin shell of the rotating wormhole against radial perturbations.
We show that the wormhole becomes more and more stable the larger its angular momentum is until the angular momentum reaches a critical value
and that the behavior of a condition for stability significantly changes when the angular momentum exceeds the critical value. 
We find that the overcritical rotating wormhole has the radius of the thin shell, which is stable regardless of the equation of state for the barotropic fluid.\\
$\qquad$\\
RUP-18-21, KEK-TH-2059, KEK-Cosmo-227
\end{abstract}

\maketitle

\section{Introduction}
A wormhole is a spacetime structure that connects two regions in our Universe or multiverse.
If the wormhole is spherically symmetric, static, and traversable in general relativity,
a null energy condition is broken at least at a throat as shown by Morris and Thorne~\cite{Morris:1988cz}.
The first example of wormholes in the Morris-Thorne class was considered 
in 1973 by Ellis~\cite{Ellis:1973yv} and Bronnikov~\cite{Bronnikov:1973fh} independently.
The Ellis-Bronnikov wormhole spacetime is the solution of Einstein equations and of a wave equation for a ghost scalar field.

The existence of wormholes in nature requires their stability. 
The instability of the Ellis-Bronnikov wormhole was reported by several 
authors~\cite{Shinkai:2002gv,Gonzalez:2008wd,Gonzalez:2008xk,Doroshkevich:2008xm,Bronnikov:2011if,Bronnikov:2012ch},
contrary to the conclusion of an earlier work~\cite{ArmendarizPicon:2002km}.
Torii and Shinkai showed that an all-dimensional Ellis-Bronnikov wormhole is unstable against perturbations with which the throat radius is changed~\cite{Torii:2013xba}.
On the other hand, in 2013, Bronnikov~\textit{et al.} found that a wormhole, which is filled with electrically charged dust with a negative energy density, 
with the same metric as the metric of the massless Ellis-Bronnikov wormhole in four dimensions~\cite{Shatskiy:2008us,Novikov:2012uj},
is stable under both spherically symmetric and axial perturbations~\cite{Bronnikov:2013coa}.
Their result shows that stability of wormholes depend on not only metrics but also the matters of the source of the metrics.~\footnote{%
The stability of a wormhole against decay via gravitational instanton tunneling was discussed by Cox~\textit{et al.}~\cite{Cox:2015pga}.
Their work also shows that the stability of wormholes depends on the matters of the source of the metrics.}

Teo pioneered a rotating traversable wormhole in four dimensions~\cite{Teo:1998dp},
but Teo did not pay attention to the stability of wormholes as much as Morris and Thorne~\cite{Morris:1988cz}.
Matos and Nu\~{n}es~\cite{Matos:2005uh} 
claimed that rotating wormholes would have a higher possibility of being stable intuitively,
and they considered a rotating wormhole with a ghost scalar field in four dimensions.
Dzhunushaliev~\textit{et al.}~\cite{Dzhunushaliev:2013jja} investigated a five-dimensional rotating wormhole with equal angular momenta 
filled with a ghost scalar field 
and discussed its stability. 
They found that the unstable mode of the five-dimensional wormhole disappears when the wormhole rotates fast
and that the wormhole has the upper bound of the angular momenta.
Their works might show that rotating wormholes have a higher possibility to be stable 
but the effect of rotation on the stability would depend on the matters supporting the wormholes.

The class of a wormhole which is composed of two parts of spacetimes joined by a thin shell~\cite{Lanczos:1922,Lanczos:1924,Israel:1966rt} 
at the wormhole throat using Darmois-Israel matching~\cite{Darmois:1927,Israel:1966rt,Poisson:2004} was suggested by Visser~\cite{Visser:1989kh,Visser:1989kg}.
The linear stability of a thin-shell wormhole against radial perturbations was studied in Refs.~\cite{Visser:1989kg,Poisson:1995sv}.
The analysis is useful to investigate the stability of wormholes 
since one can treat it with only a few assumptions with respect to the matters supporting the wormholes.
It has been applied 
to various thin-shell wormholes~\cite{Ishak:2001az,Eiroa:2003wp,Lobo:2003xd,Lobo:2005zu,Eiroa:2007qz,Eiroa:2008ky,Lemos:2008aj,%
Mazharimousavi:2010bm,Kuhfittig:2010pb,Dias:2010uh,LaCamera:2011qi,Amirabi:2011dw,Yue:2011cq,Rahaman:2011yh,%
Kuhfittig:2012gi,Banerjee:2012aja,Bochicchio:2013hd,Sharif:2013xta,Sharif:2013nka,Sharif:2013efa,Banerjee:2013qja,%
Varela:2013xua,Mazharimousavi:2014gpa,Bejarano:2014mwa,Sharif:2014bja,Kuhfittig:2014bxa,Kokubu:2014vwa,Bhar:2015dqa,%
Eid:2015pja,Kokubu:2015spa,Kuhfittig:2015hxa,Sharif:2015lod,Eid:2016axb,Ovgun:2016ijz,Dent:2016efw,Amirabi:2016dun,Sharif:2016zts,%
Ovgun:2016ujt,Eid:2017xcg,Banerjee:2016blr,Mazharimousavi:2017wrm,Eid:2017tkt,RubindeCelis:2017pzt,Richarte:2017iit,Eiroa:2017nar,%
Forghani:2018gza,Amirabi:2018ncf,Eid:2018iiu}.
Recently, the details of stability of thin-shell wormholes have been investigated.  
Nakao~\textit{et al.} considered a collision between a thin shell at a wormhole throat and another dust thin shell that falls into the throat,
and they obtained conditions in which the thin-shell wormhole persists after the collision~\cite{Nakao:2013hba}.
Akai and Nakao have investigated the nonlinear stability of a thin-shell wormhole~\cite{Akai:2017aro}. 
  
The treatment of a rotating thin shell is more difficult than a spherically symmetric thin shell because of the low symmetry of the spacetime. 
The fact, however, was shown that rotating thin shells in three dimensions 
are more tractable than ones in higher dimensions~\cite{Mann:2008rx,Crisostomo:2003xz,Ovgun:2016qiu}.
In Ref.~\cite{Mazharimousavi:2014tba}, Mazharimousavi and Halilsoy cut and pasted two parts of BTZ spacetimes~\cite{Banados:1992wn,Banados:1992gq} 
and made a rotating thin-shell wormhole in three dimensions to study the stability of the wormhole.
They claimed that one side of a throat counterrotates against the other side of the throat 
and concluded that counterrotational effects make the wormhole stable. 
However, it seems to be difficult to understand their conclusion intuitively. 

In this paper, we reexamine the stability of the rotating thin-shell wormhole in three dimensions.
We cut and paste two rotating BTZ spacetimes~\cite{Banados:1992wn,Banados:1992gq} with the Darmois-Israel matching conditions
and construct the rotating thin-shell wormhole. 
The angular momenta of both sides of the throat have the same absolute value but signs opposite each other. 
We notice that the opposite signs of the angular momenta mean
that the thin shell at the throat and the two sides of the throat must corotate. 
Then, we investigate the linear stability of the wormhole against radial perturbations.

This paper is organized as follows. 
In Sec.~II, we construct the rotating thin-shell wormhole.
In Sec.~III, In Sec. III, we consider the linear stability of the thin-shell wormhole against a radial perturbation.
In Sec.~VI, we summarize our results.
In this paper, we use the units in which the light speed and $8G$, where $G$ is Newton's constant in three dimensions, are unity 
as set in Sec.~III in Ref.~\cite{Mann:2008rx}.

\section{Construction of rotating BTZ wormhole}
We construct a rotating thin-shell wormhole by using a cut-and-paste method~\cite{Darmois:1927,Israel:1966rt,Poisson:2004}. 
We consider two BTZ spacetimes~\cite{Banados:1992wn,Banados:1992gq} with a line element given by
\begin{eqnarray}\label{eq:line_element1}
ds_\pm^2
&=& g_{\mu \nu \pm}dx^\mu_\pm dx^\nu_\pm \nonumber\\
&=&-f_\pm(r_\pm)dt_\pm^2+\frac{dr_\pm^2}{f_\pm(r_\pm)}+r_\pm^2 \left( d\varphi_\pm- \frac{J_\pm}{2r_\pm^2} dt_\pm \right)^2, \nonumber\\
\end{eqnarray}
where  $f_\pm(r_\pm)$ is defined by
\begin{equation}
f_\pm(r_\pm)\equiv -M_\pm+\frac{r_\pm^2}{l_\pm^2}+\frac{J_\pm^2}{4r_\pm^2},
\end{equation}
where $M_\pm$, $J_\pm$, and $l_\pm\equiv \sqrt{-1/\Lambda_\pm}$ are a mass parameter, an angular momentum, and  
the scale of a curvature related to a negative cosmological constant $\Lambda_\pm < 0$, 
respectively.~\footnote{A wormhole with a negative cosmological constant was considered in Ref.~\cite{Ayon-Beato:2015eca}.}
We assume $M \equiv M_+=M_- > 0$ and $l \equiv l_+=l_-$ for simplicity.
The BTZ spacetime with $lM \geq \left| J_\pm \right|$ has an event horizon at $r_\pm=r_{\pm H}$,
where $r_{\pm H}$ is given by
\begin{equation}
r_{\pm H} \equiv l\sqrt{ \frac{M}{2} \left( 1+ \sqrt{1-\frac{J_\pm^2}{M^2l^2}} \right) },
\end{equation}
while the spacetime with $lM < \left| J_\pm \right|$ has a naked singularity at $r_\pm=0$.

\begin{figure}[htbp]
\begin{center}
\includegraphics[width=87mm]{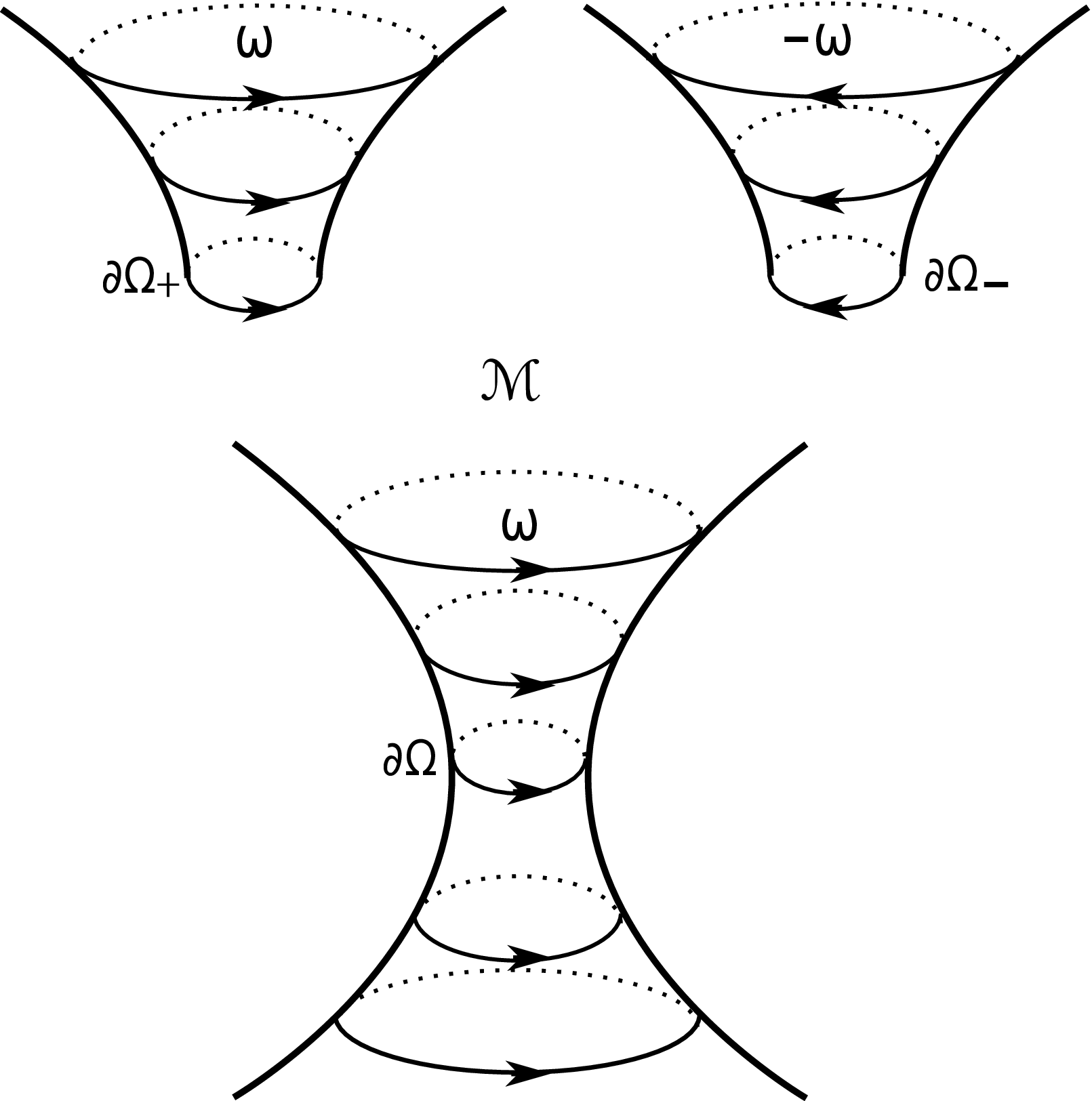}
\end{center}
\caption{Construction of a rotating wormhole with a cut-and-paste method. 
We cut two BTZ spacetimes with angular momenta $J_\pm$
and identify the boundaries of the manifolds $\partial \Omega_\pm$ that are the timelike hypersurfaces 
$\partial \Omega \equiv \partial \Omega_+ = \partial \Omega_-$,
and we obtain a manifold $\mathcal{M}$.
The second Darmois-Israel junction condition implies $J_-=-J_+$.
Thus, $\omega(a)\equiv  \omega_+(a)=-\omega_-(a)$, 
where $\omega_\pm(r_\pm)$ is defined as $\omega_\pm(r_{\pm}) \equiv -g_{t\varphi \pm}/g_{\varphi \varphi \pm}=J_\pm/2r_\pm^2$, is obtained.
}
\label{fig:rotating_WH}
\end{figure}

We consider two BTZ spacetimes and remove from them regions
\begin{equation}
\Omega_\pm \equiv \left\{ r_{\pm} < a  \:|\:  a>r_{\mathrm{b}}  \right\},
\end{equation}
where $a$ is a constant and $r_{\mathrm{b}}\equiv r_{\pm H}$ for $lM \geq \left| J_\pm \right|$ and $r_{\mathrm{b}}\equiv 0$ for $lM < \left| J_\pm \right|$.
The rest of the manifolds has the boundaries that are the timelike hypersurfaces described as 
\begin{equation}
\partial \Omega_\pm \equiv \left\{ r_{\pm} = a  \:|\:  a>r_{\mathrm{b}}  \right\}.
\end{equation}
We identify the two hypersurfaces 
\begin{equation}
\partial \Omega \equiv \partial \Omega_+ = \partial \Omega_-,   
\end{equation}
and we obtain a manifold $\mathcal{M}$ with two regions glued by a throat located at $\partial \Omega$ as shown in Fig.~\ref{fig:rotating_WH}.
 
We permit that the radius of the throat $a$ is a function of time, $a=a(t)$, since we are interested in the stability of the rotating wormhole.
By introducing an azimuth coordinate $\phi_\pm$ given by
\begin{eqnarray}
d\phi_\pm \equiv d\varphi_\pm-\frac{J_\pm dt_\pm}{2a^2(t_\pm)},
\end{eqnarray}
the line element~(\ref{eq:line_element1}) in a corotating frame on the timelike hypersurface, i.e., on the throat, is rewritten as  
\begin{eqnarray}
ds_\pm^2
&=&-f_\pm(r_\pm)dt_\pm^2+\frac{dr_\pm^2}{f_\pm(r_\pm)} \nonumber\\
&&+r_\pm^2 \left[ d\phi_\pm +\frac{J_\pm}{2}\left( \frac{1}{a^2(t_\pm)}-\frac{1}{r_\pm^2} \right)dt_\pm \right]^2.
\end{eqnarray}
We define $\left[ T \right]$ as 
\begin{equation}
\left[ T \right] \equiv \left. T_+ \right|_{\partial \Omega} - \left. T_- \right|_{\partial \Omega}
\end{equation}
for any tensorial function $T$.

From the first junction condition $[h_{ij}]=0$,
the induced metric $h_{ij}$ on the timelike hypersurface with $t_\pm=t_\pm(\tau)$ and $r_\pm=a(t_\pm(\tau))=a(\tau)$ is given by
\begin{eqnarray}
ds_{\partial \Omega}^2
=h_{ij}dy^idy^j
=-d\tau^2+a^2(\tau) d\phi^2,
\end{eqnarray}
where $\tau$ is the proper time of an observer on the timelike hypersurface
and $\dot{t}_\pm$ is obtained as
\begin{equation}\label{eq:Junc}
 \dot{t}_\pm
=\frac{\sqrt{\dot{a}^2 +f_{\pm}}}{f_\pm},
\end{equation}
where $f_\pm=f_\pm(a)$ and where the dot denotes a differentiation with respect to $\tau$. 
The induced basis vectors $e_{i \pm}^\mu$ on the timelike hypersurface are obtained as
\begin{equation}
e_{\tau \pm}^\mu \frac{\partial}{\partial x^\mu_\pm}
=\frac{\sqrt{\dot{a}^2 +f_{\pm}}}{f_\pm} \frac{\partial}{\partial t_\pm} + \dot{a} \frac{\partial}{\partial r_\pm}
\end{equation}
and
\begin{equation}
e_{\phi \pm}^\mu \frac{\partial}{\partial x_\pm^\mu}=\frac{\partial}{\partial \phi_\pm}.
\end{equation}
The 3-velocity of the observer on the hypersurface is given by $u_\pm^\mu=e_{\tau \pm}^\mu$.
The vectors $n_{\mu \pm}$ normal to the hypersurface are given by
\begin{equation}
n_{\mu \pm} dx^\mu_\pm 
= \pm \left(-\dot{a} dt_\pm+ \frac{\sqrt{\dot{a}^2 +f_\pm}}{f_\pm} dr_\pm \right).
\end{equation}

The second junction condition is given by 
\begin{equation}
\left[ K^i_j \right]=0,
\end{equation}
where $K_{ij}$ is an extrinsic curvature of the timelike hypersurface defined by
\begin{equation}\label{eq:K_ij}
K_{ij}\equiv  e^\mu_i e^\nu_j \nabla_\nu  n_\mu.
\end{equation}
The extrinsic curvature is obtained as
\begin{eqnarray}\label{eq:K_tautau}
K^\tau_{\tau \pm}&=& \pm \frac{2\ddot{a}+f'_\pm}{2\sqrt{\dot{a}^2 +f_\pm}}, \\\label{eq:K_phiphi}
K^\phi_{\phi \pm}&=& \pm \frac{\sqrt{\dot{a}^2 +f_\pm}}{a}, \\\label{eq:K_tauphi}
K^\tau_{\phi \pm}&=& \mp \frac{J_\pm}{2a},
\end{eqnarray}
where $f'_\pm=f'_\pm(a)$ is defined by
\begin{equation}
f'_\pm \equiv \left. \frac{df_\pm}{dr_\pm}\right|_{r_\pm=a}
\end{equation}
and the trace of the extrinsic curvature is obtained as
\begin{equation}\label{eq:K}
K_\pm \equiv K^i_{i \pm}= \pm \frac{2\ddot{a}+f'_\pm}{2\sqrt{\dot{a}^2 +f_\pm}} \pm \frac{\sqrt{\dot{a}^2 +f_\pm}}{a}.
\end{equation}
The $(\tau,\phi)$ component of the second junction condition 
\begin{equation}
\left[ K^\tau_\phi \right] =-\frac{J_+ + J_-}{2a}=0
\end{equation}
implies $J \equiv J_+=-J_-$.
Therefore, we obtain 
\begin{eqnarray}
f  &\equiv& f_+  = f_-  = -M+\frac{a^2}{l^2}+\frac{J^2}{4a^2} \\
f' &\equiv& f'_+ = f'_- = \frac{2a}{l^2}-\frac{J^2}{2a^3}
\end{eqnarray}
on the timelike hypersurface.
The other components of $\left[ K^i_j \right]$ given by
\begin{equation}
\left[ K^\tau_\tau \right]
=\frac{2\ddot{a}+f'}{\sqrt{\dot{a}^2 +f}}
\end{equation}
and
\begin{equation}
\left[ K^\phi_\phi \right]
=\frac{2\sqrt{\dot{a}^2 +f}}{a} 
\end{equation}
violate the second junction condition.
Thus, the rotating wormhole spacetime is singular at the throat, and a thin shell is needed there.
The thin shell at the timelike hypersurface follows the Einstein equations
\begin{equation}\label{eq:Einstein_eq}
\pi S^i_j= - \left[ K^i_j \right] + \left[ K \right] \delta^i_j,
\end{equation} 
where $S^i_j$ is the surface stress-energy tensor of the thin shell filled with a perfect fluid described by
\begin{equation}\label{eq:S_ij}
S^i_j=(\sigma+p)U^iU_j+p \delta^i_j,
\end{equation}
where $\sigma$, $p$, and $U_i$ are the surface energy density, the surface pressure, and the 2-velocity of the thin shell, respectively.
Here, the 2-velocity of the thin shell $U_i$ is given by 
$U_idy^i \equiv u_{\mu \pm} e^\mu_{i \pm} dy^i =-d\tau$, 
and $\left[ K \right]$ is given by $\left[ K \right]=\left[ K^\tau_\tau \right]+\left[ K^\phi_\phi \right]$.
The $(\tau,\tau)$ and $(\phi,\phi)$ components of the Einstein equations~(\ref{eq:Einstein_eq}) give 
\begin{equation}\label{eq:S^tau_tau}
\sigma=-S^\tau_\tau=-\frac{2\sqrt{\dot{a}^2 +f}}{\pi a}
\end{equation}
and
\begin{equation}\label{eq:S^phi_phi}
p=S^\phi_\phi= \frac{2\ddot{a}+f'}{\pi \sqrt{\dot{a}^2 +f}},
\end{equation}
respectively.
From Eqs.~(\ref{eq:S^tau_tau}) and (\ref{eq:S^phi_phi}), the equation of energy conversation is obtained as
\begin{equation}
\frac{d(2\pi a \sigma)}{d\tau} +p\frac{d (2\pi a)}{d\tau}=0.
\end{equation}

\section{Stability of the rotating wormhole}
The $(\tau,\tau)$ and $(\phi,\phi)$ components of the Einstein equations~(\ref{eq:Einstein_eq}) are rewritten as
\begin{equation}\label{eq:Einstein_eq2}
\dot{a}^2+f-\frac{\pi^2a^2\sigma^2}{4}=0
\end{equation}
and
\begin{equation}\label{eq:dot_sigma}
\dot{\sigma}=-(\sigma+p)\frac{\dot{a}}{a}.
\end{equation}

We assume that the thin shell is filled with a barotropic fluid with the surface pressure given by $p=p(\sigma)$.
The surface pressure~$p=p(\sigma)$ and Eq.~(\ref{eq:dot_sigma}) imply 
that the surface energy density can be written in the form $\sigma=\sigma(a)$.
Thus, the motion of the shell is described by  
\begin{equation}\label{eq:Einstein_eq3}
\dot{a}^2+V(a)=0,
\end{equation}
where $V(a)$ is the effective potential defined as
\begin{equation}\label{eq:V}
V(a) \equiv f(a)-\frac{\pi^2a^2\sigma^2(a)}{4}.
\end{equation}
The first and second derivatives of $V(a)$ with respect to $a$ are given by 
\begin{equation}\label{eq:V'}
V'(a) = f'(a)+\frac{\pi^2a\sigma p}{2}
\end{equation}
and 
\begin{equation}\label{eq:V''}
V''(a) = f''(a)-\frac{\pi^2}{2} \left[ p^2+(\sigma +p) \sigma \beta^2 \right],
\end{equation}
respectively, where $\beta^2$ is defined as
\begin{equation}\label{eq:beta}
\beta^2(\sigma)\equiv \frac{\partial p}{\partial \sigma},
\end{equation}
the prime is a differentiation with respective to $a$, 
and $f''(a)$ is obtained as
\begin{equation}
f''(a)=\frac{2}{l^2}+\frac{3J^2}{2a^4}.
\end{equation}

We consider a situation in which the rotating shell does not move in the radial direction. The shell is characterized by constants $a_0$, 
\begin{equation}\label{eq:S^tau_tau_st}
\sigma_0=-\frac{2\sqrt{f_0}}{\pi a_0}
\end{equation}
and
\begin{equation}\label{eq:S^phi_phi_st}
p_0=\frac{f'_0}{\pi \sqrt{f_0}},
\end{equation}
where $f_0$ and $f'_0$ are defined as $f_0 \equiv f(a_0)$ and $f'_0 \equiv f'(a_0)$, respectively.
By introducing $x\equiv a/(l\sqrt{M})$ and $x_0\equiv a_0/(l\sqrt{M})$, 
the effective potential $V(x)$ can be expanded in the power of $x-x_0$ as
\begin{equation}
V(x)=\frac{1}{2} \left. \frac{d^2V}{dx^2} \right|_{x=x_0} (x-x_0)^2 +O\left( (x-x_0)^3 \right)
\end{equation}
since $V(x_0)$ and $\left. dV/dx \right|_{x=x_0}$ vanish.
The thin shell is stable against linearized fluctuations in the radial direction for $\left. d^2V/dx^2 \right|_{x=x_0} >0$, 
while it is unstable for $\left. d^2V/dx^2 \right|_{x=x_0} <0$.
Here, $\left. d^2V/dx^2 \right|_{x=x_0}$ is obtained as 
\begin{eqnarray}
\left. \frac{d^2V}{dx^2} \right|_{x=x_0} 
&=& \frac{1}{Mx_0^4} \left[ \frac{-8x_0^6+12j^2x_0^4-6j^2x_0^2+j^4}{4x_0^4-4x_0^2+j^2} \right. \nonumber\\
&& \left. +( 2x_0^2-j^2 ) \beta_0^2 \right],
\end{eqnarray}
where $j$ and $\beta_0$ are defined as $j\equiv J/(lM)$ and $\beta_0\equiv \beta(\sigma_0)$, respectively.
The thin shell is stable if the condition 
\begin{equation}\label{eq:Stable}
\frac{-8x_0^6+12j^2x_0^4-6j^2x_0^2+j^4}{4x_0^4-4x_0^2+j^2} 
+ \left( 2x_0^2-j^2  \right) \beta_0^2 >0
\end{equation}
is satisfied.
The behaviors of inequality~(\ref{eq:Stable}) with an overcritical angular momentum $\left| j \right| > 1$
are different from the ones with a critical and a subcritical angular momentum $\left| j \right| \leq  1$.
We classify the rotating wormhole into three cases below.

\subsection{Subcritical rotating case $\left| j \right| < 1$}
\begin{figure*}[htbp]
\begin{center}
\includegraphics[width=70mm]{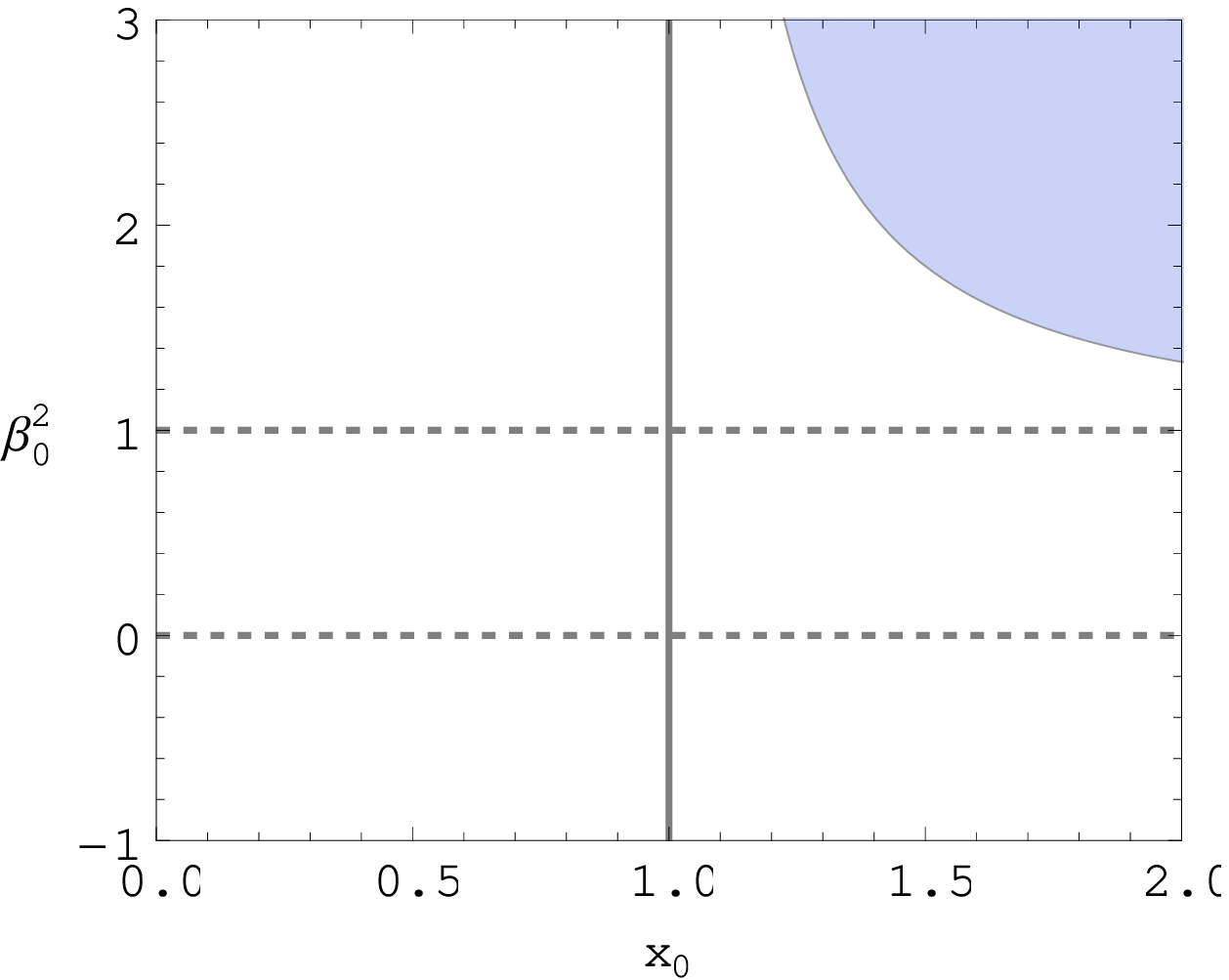}
\includegraphics[width=70mm]{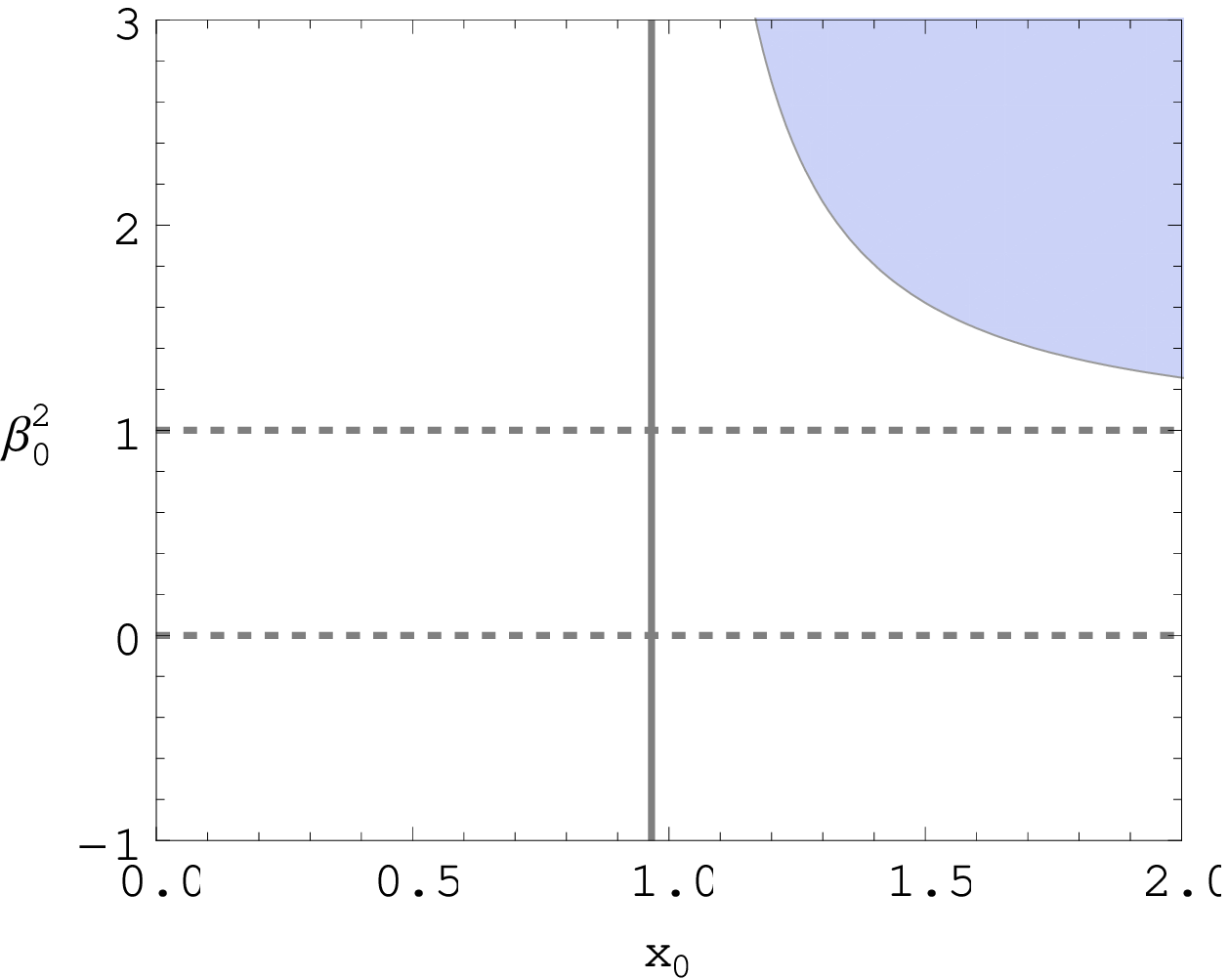}
\includegraphics[width=70mm]{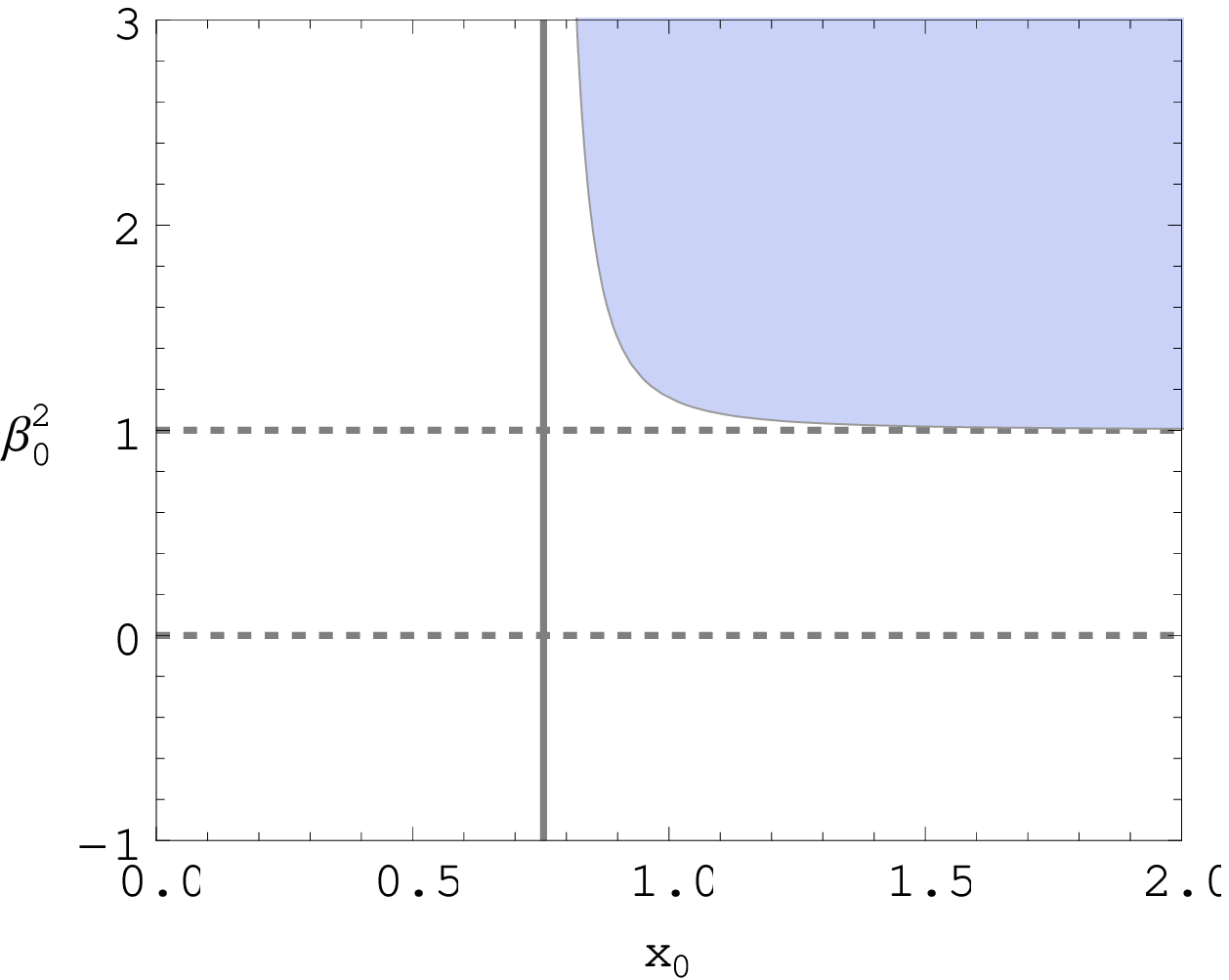}
\includegraphics[width=70mm]{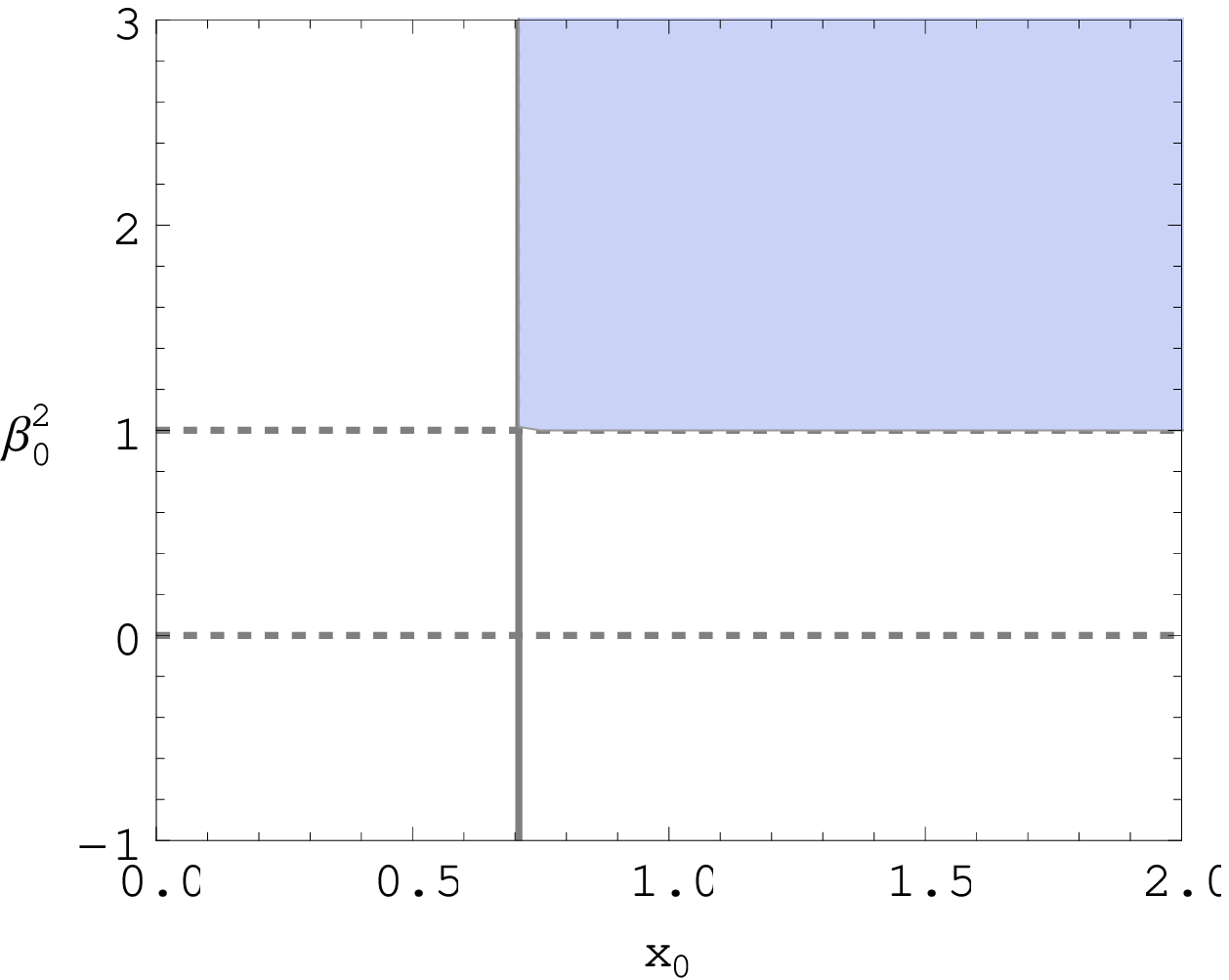}
\end{center}
    \caption{Stability of the rotating wormhole for $j=0$ (left top), $|j|=0.5$ (right top), $|j|=0.99$ (left bottom), and $|j|=1$ (right bottom). 
    Shaded regions indicate stability. Solid lines denote the radius of the event horizon for each $j$.
    One finds stable regions increase as $|j|$ increases.}
\label{fig:sub-critical-critical}
\end{figure*}

In the subcritical rotating case $\left| j \right| < 1$,
the factor $2x_0^2-j^2$ is positive 
since $x_0$ is defined in the region $x_0 > x_H$, where $x_H\equiv r_H/(l\sqrt{M})=\sqrt{\left( 1+\sqrt{1-j^2} \right)/2}$.
In this case, the stable condition~(\ref{eq:Stable}) is rewritten as 
$\beta_0^2 > A_0$,
where $A_0=A_0(x_0)$ is defined as
\begin{equation}
A_0(x_0)\equiv \frac{8x_0^6-12j^2x_0^4+6j^2x_0^2-j^4}{(2x_0^2-j^2)(4x_0^4-4x_0^2+j^2)}. 
\end{equation}
$A_0$ monotonically decreases as $x_0$ increases for $x_0 > x_H$ and 
$A_0\rightarrow 1$ in a limit $x_0 \rightarrow \infty$.
Thus, the wormhole with $\beta_0^2<1$ is unstable.
$A_0$ diverges $+\infty$ in a limit $x_0 \rightarrow x_H+0$.
This means that the rotating wormhole with any $\beta_0^2$ becomes unstable in the near-horizon limit $x_0 \rightarrow x_H+0$. 
Figure~\ref{fig:sub-critical-critical} shows stable and unstable regions. 
The more rapidly the wormhole rotates, the larger stable region is on the $x_0 \beta_0^2$ plane. 

\subsection{Critical rotating case $\left| j \right| = 1$}
In the critical rotating case $\left| j \right| = 1$,
inequality~(\ref{eq:Stable}) becomes 
$\beta_0^2>1$ and $x_0>x_H=\sqrt{1/2}$.
The wormhole with $\beta_0^2<1$ is unstable.
The stable and unstable regions are shown in Fig.~\ref{fig:sub-critical-critical}.

\subsection{Overcritical rotating case $\left| j \right| > 1$}
\begin{figure*}[htbp]
\begin{center}
\includegraphics[width=70mm]{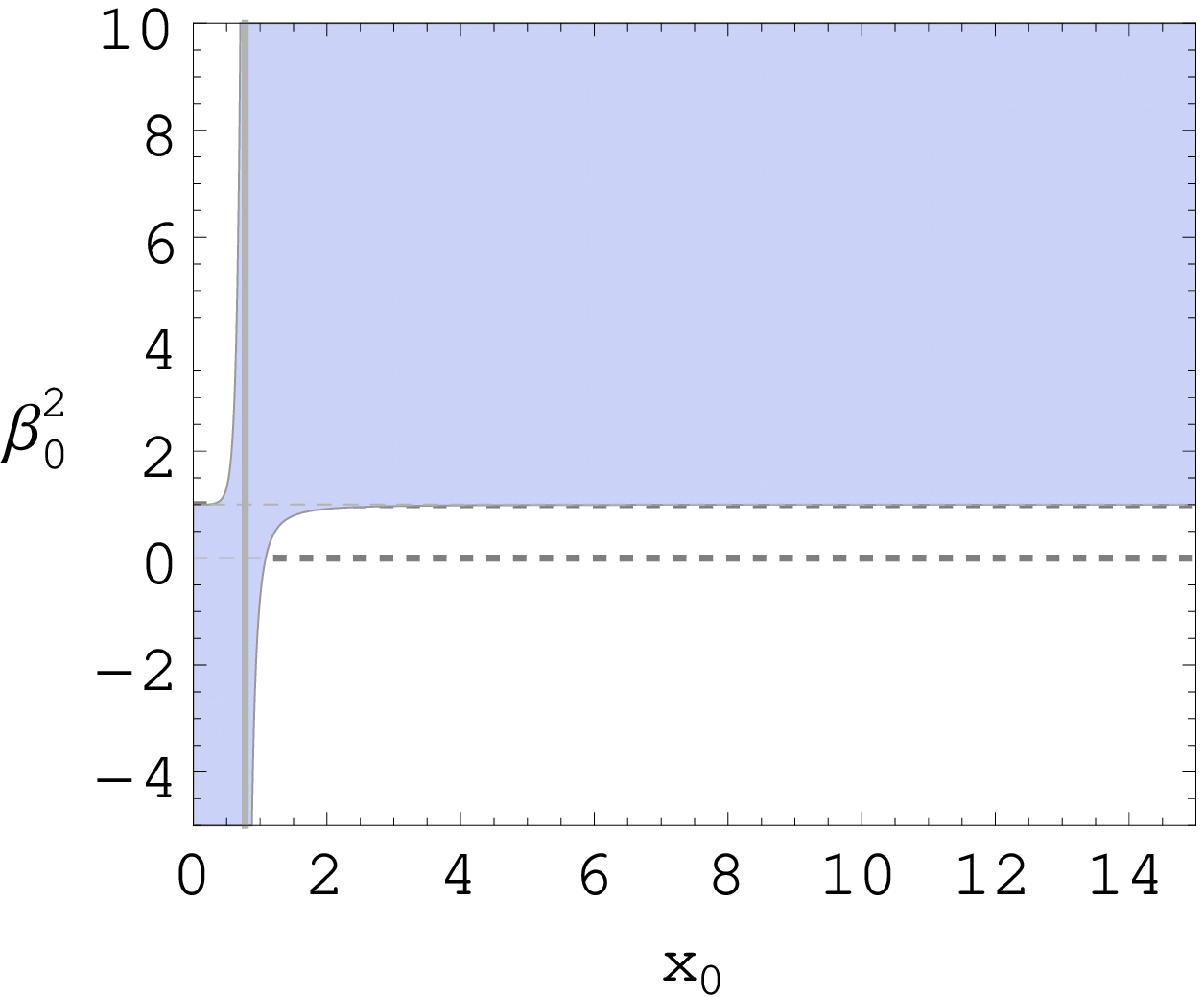}
\includegraphics[width=70mm]{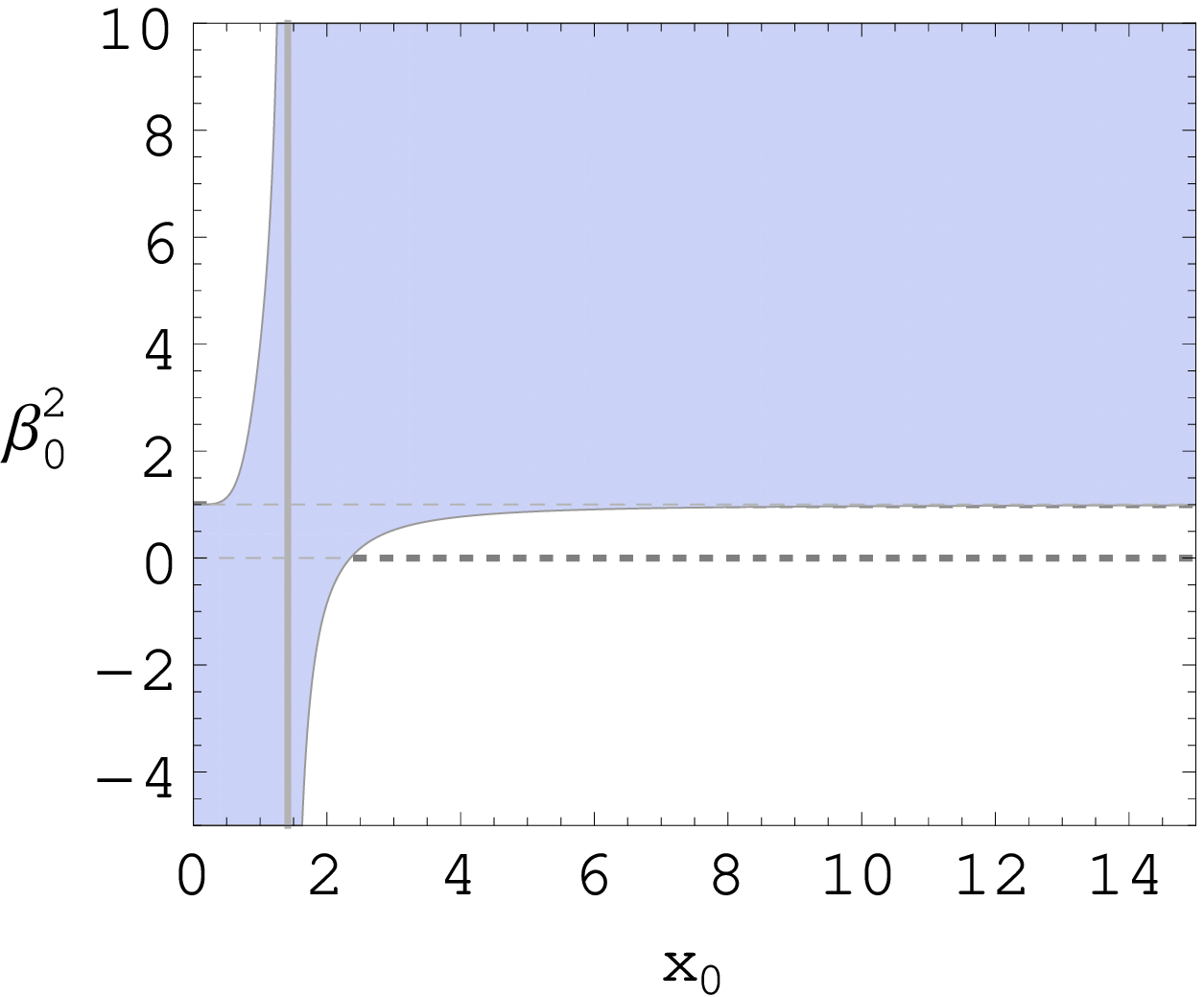}
\includegraphics[width=70mm]{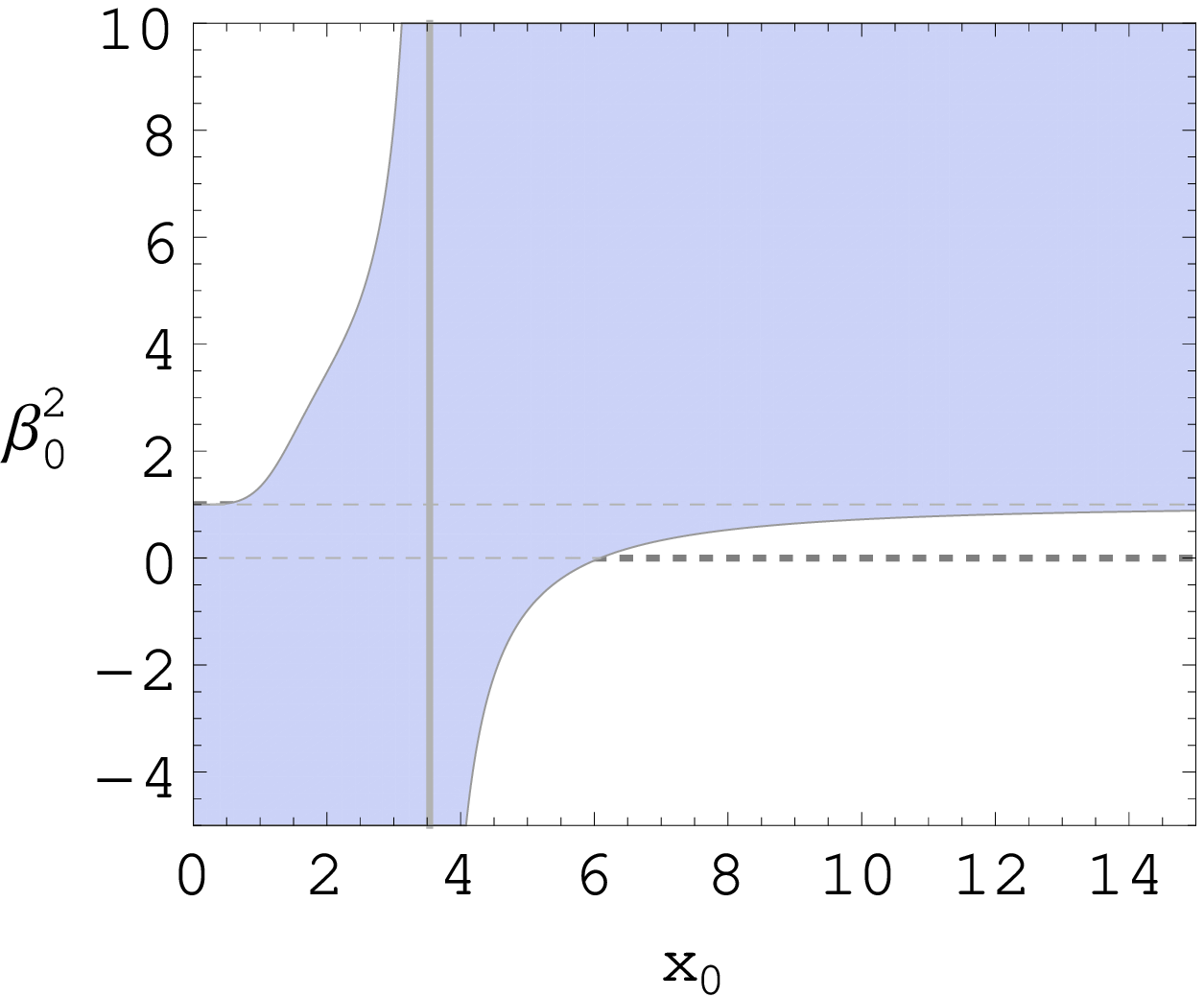}
\includegraphics[width=70mm]{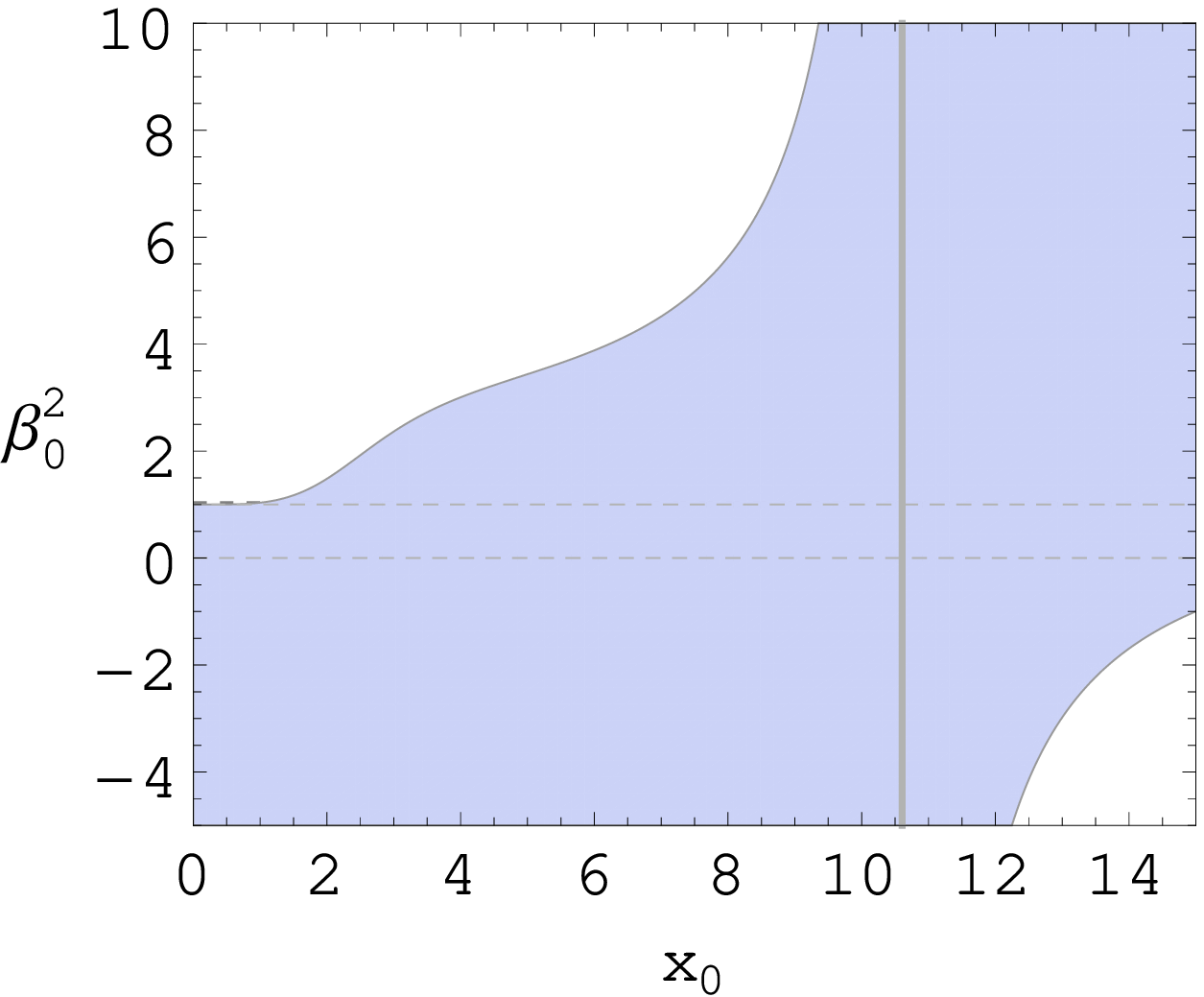}
\end{center}
\caption{Stability regions for the rotating wormhole with $|j|=1.1$ (left top), $|j|=2$ (right top), $|j|=5$ (left bottom), and $|j|=15$ (right bottom). 
Shaded regions indicate stability. Solid lines represent $x_0=|j|/\sqrt{2}$.}
\label{fig:over-critical}
\end{figure*}

In the overcritical rotating case $\left| j \right| > 1$,
the stability condition~(\ref{eq:Stable}) is given by $\beta_0^2 < A_0$
($\beta_0^2 > A_0$) 
for $0<x_0<\left| j \right|/\sqrt{2}$ ($\left| j \right|/\sqrt{2}<x_0$), and 
$A_0$ approaches $\infty$ ($-\infty$) in a limit $x_0 \rightarrow \left| j \right|/\sqrt{2}-0$ ($\left| j \right|/\sqrt{2}+0$).
$A_0$ monotonically increases as $x_0$ increases in both regions $0<x_0<\left| j \right|/\sqrt{2}$ and $\left| j \right|/\sqrt{2}<x_0$.
Note $A_0\rightarrow 1$ in both the limits $x_0 \rightarrow \infty$ and $x_0 \rightarrow +0$. 
Figure~\ref{fig:over-critical} shows stable and unstable regions. 
We notice that the rotating wormhole with any $\beta_0^2$, i.e., any barotropic fluid, is stable when the throat is at $x_0=\left| j \right|/\sqrt{2}$.
Roughly speaking, the rapidly rotating wormhole is stable if the throat is near $x_0=\left| j \right|/\sqrt{2}$ 
on the $x_0 \beta_0^2$ plane. 
We also notice that a stable region with $0<\beta_0^2<1$ exists on the $x_0 \beta_0^2$ plane.

\section{Discussion and Conclusion}
We have cut and pasted two BTZ spacetimes, and we have constructed a rotating thin-shell wormhole by a Darmois-Israel method.
The second Darmois-Israel junction condition imposes that two BTZ spacetimes should have an equal absolute value of angular momenta but signs opposite each other. 
We have shown that the wormhole composed of a thin shell and both sides of a throat corotate as shown in Fig.~\ref{fig:rotating_WH}, 
while Ref.~\cite{Mazharimousavi:2014tba} concluded that a side of the throat counterrotates against the other side.

We have investigated the linear stability of the wormhole against a radial perturbation.
We have concentrated on the thin shell filled with a barotropic fluid.
The wormhole has a critical angular momentum  $\left| j \right| = 1$,
and we find the different behaviors of the stability on the $x_0 \beta_0^2$ plane 
in the following two cases.

In the subcritical and critical rotating cases $\left| j \right| < 1$,
the more rapidly the wormhole rotates, the larger stable region is on a $x_0 \beta_0^2$ plane
as shown Fig.~\ref{fig:sub-critical-critical}.
In these cases, the wormhole with $\beta_0^2<1$ is unstable.
In a limit at which the throat approaches the event horizon $x_0 \rightarrow x_H+0$,
the surface pressure $p$~(\ref{eq:S^phi_phi_st}) diverges, 
and the wormhole with any $\beta_0^2$ is unstable on the $x_0 \beta_0^2$ plane.

In the overcritical rotating case $\left| j \right| > 1$,
we notice that the wormhole with the throat at a radius $x_0=\left| j \right|/\sqrt{2}$
is stable without dependence on $\beta_0^2$, i.e., the equation of state for the barotropic fluid
~\footnote{We notice that a charged thin-shell wormhole is also stable with any $\beta_0^2$ when some conditions are satisfied~\cite{Eiroa:2003wp}.}
and that the stable region of the wormhole with $0<\beta_0^2<1$ exists on the $x_0 \beta_0^2$ plane.
In this case, one cannot say that the wormhole becomes more stable the faster wormhole rotates.

We have considered the rotating wormhole which is composed of two parts of two BTZ spacetimes just because we can treat it easily.
The simplicity would help us in the further development of the stability of rotating thin-shell wormholes
such as the collision of thin shells~\cite{Nakao:2013hba} and nonlinear stability~\cite{Akai:2017aro}.
We hope that this article will stimulate further work in this direction.

In the rest of this section, we comment on rotating wormholes.
In Refs.~\cite{Kleihaus:2014dla,Chew:2016epf}, 
a rotating wormhole solution with a ghost scalar field in four dimensions was obtained, 
and the maximum of the angular momentum of the rotating wormhole was found.
The metric of the maximal rotating wormhole is correspondent with the one of a maximal rotating Kerr black hole.
A rotating wormhole solution with an equal angular momenta filled with a ghost scalar field in five dimensions also has 
a maximal angular momentum~\cite{Dzhunushaliev:2013jja}. 
The metric of the maximal rotating wormhole is correspondent with the one of a maximal rotating Myers-Perry black hole~\cite{Myers:1986un,Myers:2011yc} 
in five dimensions.
These facts do not mean that any wormholes cannot rotate rapidly 
since the upper bound of the angular momentum would depend on matters that support wormholes. 

It was shown that the center-of-mass
energy of two falling particles can be very large if the
particles collide at the throat of a very highly rotating
wormhole~\cite{Tsukamoto:2014swa,Zaslavskii:2015vaa}. 
This implies that very highly
rotating wormholes would be unstable against particle
collisions.

\section*{Acknowledgements}
The authors thank S.~Hou, K.-i.~Nakao, and A.~Naruko for their useful comments.

\end{document}